\documentclass[12pt]{article}
\usepackage[utf8]{inputenc}
\usepackage[top=0.7in,bottom=1in,left=1.1in,right=1.1in,includehead]{geometry}
\usepackage{epsfig,verbatim,cite}
\usepackage{hyperref}
\usepackage{mathtools}
\usepackage[usenames,dvipsnames,svgnames,table]{xcolor}
\usepackage{epsfig,amsmath,amsfonts,amssymb,arydshln}

\numberwithin{equation}{section}

\def\a{\alpha} \def\b{\beta} \def\g{\gamma}  \def\e{\epsilon}
  \def\h{\eta} \def\q{\theta}
  \def\k{\kappa} \def\l{\lambda} \def\m{\mu}
\def\n{\nu} \def\x{\xi}   \def\r{\rho}
 \def\s{\sigma}   \def\f{\varphi}

\def\D{\Delta} 
   
   \def\L{\Lambda} 
   
 \def\S{\Sigma}  
\def\F{\Phi}

\def\bM{\bar{M}}\def\bN{\bar{N}}

\def\fr{\frac}  \def\dt{\partial}

\def\ph{\phantom}
\def\mc{\mathcal}
\def\mA{\mathcal{A}}

\def\mE{\mathcal{E}}
\def\mF{\mathcal{F}}

\def\mH{\mathcal{H}}
\def\mK{\mathcal{K}}
\def\mL{\mathcal{L}}
\def\mM{\mathcal{M}}
\def\mN{\mathcal{N}}

\def\mZ{\mathcal{Z}}

\def\tx{\tilde{x}}

\def\hA{\hat{A}}

\def\hM{\hat{M}}
\def\hN{\hat{N}}

\def\tdt{\tilde{\partial}}

\def\tm{\times}

\def\XX{\mathbb{X}}

\newcommand\bqa {\begin{eqnarray}}
\newcommand\eqa {\end{eqnarray}}

\newcommand{\bear}{\begin{array}}
\newcommand{\enar}{\end{array}}

\newcommand{\be}{\begin{equation}}
\newcommand{\ee}{\end{equation}}
\def\bea{\begin{eqnarray}}
\def\eea{\end{eqnarray}}

\def\DD{{\mathcal{D}}}

\begin{document}

\begin{titlepage}
\ph{preprint}

\vfill

\begin{center}
   \baselineskip=16pt
   {\large \bf 
Gauge field fluxes and Bianchi identities in extended field theories 
   }
   \vskip 2cm
    Edvard T. Musaev$^\dagger{}^\bullet$\footnote{\tt musaev.et@phystech.edu}
       \vskip .6cm
             \begin{small}
                          {\it
                          $^\dagger$Moscow Institute of Physics and Technology,                         Laboratory of High Energy Physics\\
                          Institutskii per., 9, 141702, Dolgoprudny, Russia\\[0.5cm] 
                           $^\bullet$Kazan Federal University, Institute of Physics\\
                          Kremlevskaya 16a, 420111, Kazan, Russia                         
                          } \\ 
\end{small}
\end{center}

\vfill 
\begin{center} 
\textbf{Abstract}
\end{center} 
\begin{quote}
Tensor hierarchy of Exceptional Field Theories contains gauge fields satisfying certain Bianchi identities with source part governing interaction with standard and exotic branes. These are responsible for tadpole cancellation  in compactification schemes and provide consistency constraints for cosmological model building. Analysis of reduction of (10+10)-dimensional DFT into (D+d+d)-dimensional split DFT allows to consider all Bianchi identities of the theory in the form analogous to the ExFT approach. Here we review in details and elaborate on these ideas. \\ Contribution to proceedings of the VI International Conference ``Models of Quantum Field Theory''.
\end{quote} 
\vfill
\setcounter{footnote}{0}
\end{titlepage}

\newpage
\setcounter{page}{2}


\section{Introduction}

One of the major goals of string theory as a consistent formulation of quantum gravity is description of inflationary cosmological models in four dimensions. To comply with observational data one considers space-time configurations where the four dimensional space-time of de Sitter type is supplemented by a compact 6d manifold whose size is small enough to be hidden from direct searches so far. Fortunately, equations of motion of supergravity have solutions of precisely the required form, of which the simplest case is the 6d torus. However, toroidal backgrounds without fluxes cannot reproduce any realistic phenomenology, since these preserve to much supersymmetry, cannot provide scalar potential with stable minima and hence generate non-chiral effective theories with too many massless scalar fields. The problem of finding a better option for the internal compact manifold which provides masses and charges for scalar fields of the effective 4d theory and generate a suitable inflaton potential is usually addressed as the problem of moduli stabilization. Finding an appropriate compactification scheme for string theory including fluxes and branes has been intensively investigated (for review see \cite{Grana:2005jc,Blumenhagen:2006ci,Cline:2006hu}). The most famous example of cosmological model based on compactifications with D-branes is the KKLT scenario \cite{Kachru:2003aw}(see also \cite{Kallosh:2018psh} and references therein for the recent discussion on consistency of the model).

To illustrate the need of fluxes and non-trivial geometry to provide scalar masses let us consider a toy model of a 1+5-dimensional theory of gravity interacting with electromagnetic field and perform dimensional reduction down to 4d as in \cite{Denef:2007pq}. The internal 2d manifolds can all be classified by the number of handles and we will consider volume of the internal manifold as the only scalar modulus of the resulting theory. Hence, one starts with the theory
\begin{equation}
S=\int d^6 x \sqrt{-G}\Big(M_6^4 {\cal R}_{(6)}[G]-M_6^2 F^2\Big),
\end{equation}
where ${\cal R}_{(6)}[G]$ is Riemann curvature of the metric $G$ in the full 1+5-dimensional space-time, $F^2$ is the contribution from the electromagnetic field and $M_6$ is the corresponding Planck mass which is there for dimensional reasons. Consider the usual compactification ansatz for the metric
\begin{equation}
ds^2=e^{}g_{\m\n}dx^\m dx^\n+R(x)^2 h_{mn}dy^m dy^n,
\end{equation}
where $\{x^\m\}$ and $\{y^m\}$ are coordinates on the external and internal space respectively, $h_{mn}$ is metric of unit volume on the internal space and  the field $R(x)$ corresponds to the volume modulus in the theory. For simplicity we assume that there are no $G_{\m m}$ components in the full metric. To keep the kinetic term in the canonical form one rescales the external metric and the final action becomes
\begin{equation}
S_{eff}=\int d^4x \Big({\cal R}_{(4)}[h]+h^{mn}\dt_m R \dt_n R + V(R) +\dots \Big),
\end{equation}
where the potential $V(R)$ depends on geometry of the internal manifold and on flux of the gauge field. Ellipses denote contributions of higher order in perturbation theory and interactions which are of no interest for the discussion. The potential can be written as 
\begin{equation}
\begin{aligned}
V(R) \sim &\ (2g-2)\fr{1}{R(x)^4}+\fr{n^2}{R(x)^6},\\
n = &\ \int_{\S} F, 
\end{aligned}
\end{equation}
where $n$ is flux of the field $F_{mn}$ integrated over the internal manifold $\S$ and $g$ is genus of $\S$. These are topological invariants and hence are initial parameters of the model. 

Stability of the theory under small variations of the field $R(x)$ around a minimum of the potential crucially depend on the chosen parameters, genus and flux. In Table \ref{tab_pots} essentially different case are listed. One notices that for toroidal compactification $g=1$ with no flux $n=0$ the potential is flat $V=0$ and minimum is not represented by a single point. 
\begin{table}[h]
\centering
\label{tab_pots}
\begin{tabular}[ht]{l|c|cc}
       & $n=0$ & $n\neq 0$ \\
       \hline
$g=0$ (sphere) &  inconsistent &    Freund-Rubin (stable)           \\ 
$g=1$  (torus) &  flat &      runaway                 \\ 
$g>1$  &  runaway &            runaway      
\end{tabular}
\caption{Potential behavior for different genera $g$ and number of fluxes $n$.}
\end{table}
Hence, for each flat direction in the scalar potential the effective four-dimensional theory will contain a massless field. Compactification on  a sphere $g=0$ with no flux will result in a potential with minimum and $R(x)=0$, hence, dynamically the theory will tend to have no internal directions at all. This contradicts to our initial assumption that size of the internal space is small but finite. For higher genera $g>1$ and not fluxes one observes the runaway behavior when the potential is minimized as $R(x)\to \infty$. This corresponds to spontaneous decompactification of the theory and going back to the 1+5-dimensional space. As follows from the Table \ref{tab_pots} turning on fluxes and choosing a sphere as the internal manifold allows to stabilize the field $R(x)$ at a non-zero value at the minimum of the potential. Such class of solutions in supergravity is called Freund-Rubin solutions \cite{Freund:1980xh} and does not provide proper background for cosmological model building as the potential at the minimum is negative. However, this example illustrates how presence of fluxes and choice of geometry affects the lower-dimensional effective theory.

To provide support for gauge field on the compact manifold one must include branes into compactification scheme. To do so consistently tadpole cancellation conditions must be satisfied. These originate from an analogue of the Gauss theorem for the compact internal manifold. Indeed consider a RR $p+1$-form field $C_{p+1}$ with field strength $F_{p+2}=dC_{p+1}$ whose action reads
\begin{equation}
S=-\fr14\int d^{10}x \ {*F}_{p+1}\wedge F_{p+1}+\m \sum_{a}\int_{\S_a} d^{p+1}\x \, C_{p+1},
\end{equation}
where $\S$ denotes world-volume of the corresponding Dp-branes labeled by index $a$. Introducing a current $J^a_{p+1}$ for each of the D-branes equations of motion will take the following form
\begin{equation}
d*dC_{p+1}=\m \sum_a *J^a_{p+1}.
\end{equation}
If the sources and the fluxes are localized only on the compact manifold then the LHS above vanishes upon integrating over this manifold, and the RHS gives the total charge. This implies that the total charge of any RR field localized on a compact manifold must be zero. This is usually addresses as the tadpole cancellation condition, which can be fulfilled by introducing orientifold planes into the model (see \cite{Blumenhagen:2006ci} for review). 

Magnetic dual of a Dp-brane interacts with RR field $\tilde{C}_{7-p}$ which is  the Poincar\'e dual of the RR field $C_{p+1}$. Equation of motion for the dual field follows from the Bianchi identities for the field strength of the field $C_{p+1}$
\begin{equation}
d F_{p+2}+\cdots=d\ {*F}_{8-p}+\cdots=\m \sum_ a *J^a_{7-p},
\end{equation}
where ellipses denote possible contributions from other gauge potentials. The above equation implies that in presence of branes magnetically charged w.r.t. a gauge potential $C_p$ the corresponding field strength will be topologically non-trivial and have non-trivial Bianchi identities.  For D-branes one is able to consider only electrically charged objects and all Bianchi identities can be set trivial, however in the NS-NS sector one finds NS5 branes, which are magnetically charged w.r.t. the Kalb-Ramond 2-form field $B_2$. This field electrically couples to the fundamental string F1. 

In compactification schemes involving NS 5-branes tadpole cancellation conditions cannot be fulfilled by adding Op-planes into the model, however, one may equivalently impose vanishing of the RHS of the Bianchi identity. Based on T-duality transformations of the Gukov-Vafa-Witten superpotential and of Bianchi identities for (constant) fluxes such cancellations have been analysed in \cite{Shelton:2005cf,Aldazabal:2006up,Lombardo:2016swq}. This analysis includes not only the geometric fluxes of the NS5-brane and the KK5 monopole, but also non-geometric Q-fluxes and R-fluxes sourced by $5_2^2$ and $5_2^3$ branes respectively. Such branes interact electrically with mixed-symmetry gauge potentials $B_{(8,2)}$ and $B_{(9,3)}$, which are magnetic duals of the bivector field $\b^{mn}$. In the works \cite{Bergshoeff:2011ee,Bergshoeff:2012ex,Chatzistavrakidis:2013jqa,Blair:2017hhy} a worldvolume DBI action for such branes has been presented and coupling to the mixed-symmetry potentials has been analysed.

Understood as proper gauge fields such mixed symmetry potentials are expected to generate field strengths which must satisfy certain Bianchi identities alongside with equations of motion. In this letter we show that Double Field Theory formulation of supergravity allows to write such Bianchi identities in a T-duality covariant form, and gives a hint for M-theory generalization of the result. In Section \ref{sec:nongeom} we briefly describe the DFT approach to non-geometric fluxes in terms of Scherk-Shwarz reduction. In Section \ref{sec:split} we review the  split-form DFT as obtained from the full O(10,10) theory. In Section \ref{sec:bianchi} we describe Bianchi identities for generalised fluxes of the O(d,d) theory obtained by reduction from the full DFT and interpret the identities in terms of 5-brane sources of various orientations. Finally, in Section \ref{sec:m} we discuss extension of the described results to the case of exceptional field theory and gauge potentials of non-geometric branes of M-theory.

\section{Fluxes in Double Field Theory}
\label{sec:nongeom}

Double Field Theory provides a natural framework for addressing properties of non-geometric fluxes and the corresponding mixed-symmetry potential. This approach has been developed mainly in the works \cite{Hohm:2010jy,Hohm:2010pp,Hohm:2013nja} and review of the formalism can be found in \cite{Berman:2013eva,Hohm:2013bwa,Aldazabal:2013sca}. The approach is based on extending the space-time by coordinates corresponding to winding modes of the string and rewriting the field content in a T-duality-covariant form. For the fully doubled 10+10-dimensional space-time parametrized by coordinates $\XX^M=(x^\m,\tx_\m)$ the field content is encoded in the so-called generalised metric
\begin{equation}
\mH_{MN}=
\begin{bmatrix}
G_{\m\n}+B_{\m\k}G^{\k\l}B_{\l\n} & B_\m{}^\s \\
B_\n{}^\r & G^{\r\s}
\end{bmatrix} \quad \in \fr{O(10,10)}{O(1,9)\times O(1,9)},
\end{equation}
and the invariant dilaton $d=\f - 1/4 \log \det G$.

Dynamics of the theory is given by the action first presented in \cite{Hohm:2010pp}, which takes the following form
\begin{equation}
\label{Odd_action}
\begin{aligned}
    S=\int d^{20}\XX e^{-2d}&\left(\fr18\mc{H}^{MN}\dt_M\mc{H}^{KL}\dt_{N}\mc{H}_{KL} -\fr12
\mc{H}^{KL}\dt_L\mc{H}^{MN}\dt_N\mc{H}_{KM} - \right.\\
      & \left.- 2\dt_M d\dt_N\mc{H}^{MN}+4\mc{H}^{MN}\dt_Md\dt_Nd\lefteqn{\ph{\fr12}}\right) \, .
      \end{aligned}
\end{equation}
This action is invariant under global $O(10,10)$ transformations and under local transformations governed by generalised Lie derivative defined as
\begin{equation}
\mL_\L V^M=\L^N \dt_N V^M-(\dt_N \L^M-\dt^M \L_N) V^N.
\end{equation}
Indices are raised and lowered by the invariant tensor of $O(10,10)$
\begin{equation}
\h_{MN}=\begin{bmatrix}
0 & \mathbf{1}\\
\mathbf{1} & 0
\end{bmatrix}.
\end{equation}
For consistency of algebra of generalised Lie derivatives one must impose constraints on all fields on which it is realized \cite{Berman:2012vc}. This constraint is called section condition and for DFT can be written as
\begin{equation}
\h^{MN}\dt_M \bullet \dt_N \bullet =0,
\end{equation}
where bullets stand for any fields and their  combinations. Effectively, this boils down to the condition that all fields can depend only on a half of the total number of coordinates. The most natural choice is $\tdt^m\bullet =0$, i.e. all fields depend only on geometric coordinates $x^m$. Upon this constraint the generalised Lie derivative splits into diffeomorphisms and gauge transformations, and the action $S_{HHZ}$ reproduces the normal action of (the bosonic part of) 10-dimensional supergravity. Choosing different subsets of the total set of coordinates to be dropped is equivalent to choosing a T-duality frame. 

Based on the progress made in \cite{Jensen:2011jna,Berman:2014jsa} in the work \cite{Bakhmatov:2016kfn} it has been shown that DFT allows solutions which preserve the section constraint, but fail to satisfy equations of motion of normal supergravity as they depend on dual coordinates. These correspond to backgrounds of exotic 5-branes of the NS sector: KK vortex and Q- and R-monopoles. Backgrounds sourcedby such branes can be characterized by fluxes which are encoded in the generalised torsion $\mF_{AB}{}^C$ of DFT defined in terms of the generalised vielbein $E_M{}^A$ as
\begin{equation}
\label{gen_flux}
[E_A,E_B]_{\mc C}=\mF_{AB}{}^CE_C.
\end{equation}
The generalised vielbein is defined in the usual way as $\mH_{MN}=E_M{}^AE_N{}^B\mH_{AB}$ with diagonal and constant $\mH_{AB}$. NS fluxes in terms of components of the generalised torsion read
\begin{equation}
\begin{aligned}
H_{abc}=\mF_{abc}&& f_{ab}{}^c=\mF_{ab}{}^c, && Q_a{}^{bc}=\mF_a{}^{bc}, && R^{abc}=\mF^{abc}. 
\end{aligned}
\end{equation}
From the equation \eqref{gen_flux} one obtains components of the generalised torsion written in terms of the vielbein and its derivatives
\begin{equation}
\mF_{ABC}=-3 E_{[A}{}^ME_B{}^N\dt_M E_{C] N},
\end{equation}
and flat DFT indices are raised and lowered by the corresponding invariant tensor $\h_{AB}$. In addition one has flux corresponding to the dilaton field
\begin{equation}
\mF_A=\dt_M E_A{}^M+2E_A{}^M\dt_M d.
\end{equation}
Since these encode the same degrees of freedom as the generalised metric and the invariant dilaton, the initial action $S_{HHZ}$ can be completely rewritten in terms of $\mF_{ABC}$ and $\mF_A$. Such flux formulation of DFT has been presented in \cite{Geissbuhler:2011mx} and the Lagrangian takes the following form
\begin{equation}
\begin{aligned}
S =\int d \XX e^{-2d} \bigg(&- \fr14 \mF_{AD}{}^C \mF_{BC}{}^D \mH^{AB} - \fr{1}{12}\mF_{AC}{}^E \mF_{BD}{}^F \mH^{AB} \mH^{CD} \mH_{EF} \\&+ \mF_A\mF_B \mH^{AB}  -\fr{1}{6}\mF^{ABC}\mF_{ABC}- \mF^A\mF_A \bigg).
\end{aligned}
\end{equation}
This must be preserve generalised diffeomorphism invariance and local gauge transformations of the vielbein. These conditions imply the following constraints on the fluxes
\begin{equation}
\label{BIO1010}
\begin{aligned}
E^M_{[A}\dt_M\mF_{BCD]}-\fr34\mF^E{}_{[AB}\mF_{|E|CD]}& \equiv \mZ_{ABCD}=0,\\
E_M{}^C\dt^M {\mF}_{CAB} +2E_{[A}{}^M\dt_M{\cal F}_{B]}-{\cal F}^C {\cal F}_{CAB}&\equiv {\cal Z}_{AB}=0,\\
E_M{}^A\dt_M{\cal F}_A-\frac{1}{2}{\cal F}^A{\cal F}_A +\frac{1}{12}{\cal F}^{ABC}{\cal F}_{ABC}&\equiv \mZ=0.
\end{aligned}
\end{equation}
Solving these Bianchi identities one is able to recover the fields $\mF_{ABC}$ and $\mF_A$ in terms of the generalised vielbein as above. 

The relation between the generalised vielbein and the generalised torsion is of the same nature as the relation between the gauge field $B_{\m\n}$ and its field strength. This can be seen explicitly for the H-flux components $\mF_{\m\n\r}$ of the torsion $\mF_{MNK}$ written in curved indices as
\begin{equation}
\mF_{MNK}=E_M{}^AE_N{}^BE_{K}{}^C\mF_{ABC}.
\end{equation}
For these components we have
\begin{equation}
\mF_{\m\n\r}=3\dt_{[\m}B_{\n\r]}+\cdots,
\end{equation}
where ellipses denote terms non-linear in $B$ and which contain metric $G_{\m\n}$. Bianchi identities for $\mZ_{MNKL}$ then imply for the H-flux
\begin{equation}
\dt_{[\m}H_{\n\r\s]}+\cdots =0,
\end{equation}
where the section constraint $\tdt^\m=0$ has been imposed. For the Poincar\'e dual of the 3-form field strength $H_{(7)}=*H_{(3)}$ the above can be read as equations of motion
\begin{equation}
\nabla^\m H_{\m \n_1\dots \n_6}+\cdots=0,
\end{equation}
and must be supplemented by a proper source on the RHS. In the case in question this is the NS5-brane, which is magnetic dual of the fundamental string F1 and interacts with the Kalb-Ramond field $B_{\m\n}$ magnetically (see e.g. \cite{Townsend:1997wg,Obers:1998fb}). Hence, it interacts with the 6-form field $B_{\m_1\dots \m_6}$ electrically and the corresponding source contribution can be written as
\begin{equation}
\nabla^\m H_{\m \n_1\dots \n_6}+\cdots=j^{(0)}_{\n_1\dots \n_6}.
\end{equation}
Since the equations of motion for the magnetic potential $B_{(6)}$ are a rewriting of the Bianchi identities for the electric potential $B_{(2)}$, the latter also must be supplemented by the same source contribution
\begin{equation}
\dt_{[\m}H_{\n\r\s]}+\dots = (*j^{(0)})_{\m\n\r\s}.
\end{equation}
The same arguments can be repeated for each component of the fluxes $\mF_{MNK}$ and $\mF_M$ and the result boils down to having a T-duality covariant source term on the RHS of the Bianchi identities of DFT
\begin{equation}
\dt_{[M}\mF_{NKL]}+\fr34\mF_{MN}{}^P\mF_{KL]P}=T_{MNKL}.
\end{equation}
Bianchi identities for exotic NS fluxes of supergravity sourced by exotic 5-branes have been analysed in \cite{Chatzistavrakidis:2013jqa,Andriot:2014uda} for backgrounds of the conventional supergravity and in \cite{Bakhmatov:2016kfn,Blair:2017hhy} for backgrounds of DFT.

Bianchi identities when understood as conditions on T-duality covariant field strengths allow to introduce the corresponding dual potentials as Lagrange multipliers in the full DFT action \cite{Bergshoeff:2011mh,Bergshoeff:2016ncb}
\begin{equation}
S_{Full}=S_{HHZ}+\int d^{20}\XX e^{-2d}\Big(\mZ_{MNKL}D^{MNKL}+\mZ_{MN}D^{MN}+\mZ D\Big).
\end{equation}
The potentials $D^{MNKL}$ contain Poincar\'e dual of the 6-form $B_6$ which is magnetic partner of the Kalb-Ramond 2-form $B_{\m\n}$
\begin{equation}
D^{\m_1\dots \m_4}=\e^{\m_1\dots \m_{10}}B_{\m_5\dots \m_{10}}.
\end{equation}
Other components contain dual graviton and potentials interacting with exotic branes, for which reason the theory cannot be written purely in terms of the dual potentials at the fully non-linear level. Linearized version of the dual theory has been presented in \cite{Bergshoeff:2016ncb,Bergshoeff:2016gub} and gauge transformations have been analysed.

When considering the split-form of DFT the potentials $D^{MNKL}, D^{MN}, D$ drop into $p$-forms with values in tensors of various rank of the remaining $O(d,d)$ symmetry. These can be interpreted as potentials interacting with NS 5-branes differently embedded into the partially doubled space-time. As it has been found in \cite{Bergshoeff:2011zk} NS 5-branes in $D$ dimensions interact with the following magnetic gauge potentials
\begin{equation}
\label{pot_D}
\begin{aligned}
&D_{D-4}, && D_{D-3,M},&& D_{D-2,MN}, && D_{D-1,MNK},   && D_{D,MNKL},  \\
& &&  && D_{D-2}, &&  D_{D-1,M},  && D_{D,MN},\\
& &&  &&  && && D_{D},
\end{aligned}
\end{equation}
where the $O(d,d)$ indices are understood to be totally antisymmetric. For $D=0$ corresponding to the full DFT only the last column survives returning us back to the previous case. Given the relation between potentials and branes and the fact that the split-form DFT has the same structure as exceptional field theories, it is natural to ask what are the corresponding Bianchi identities for the potentials above and what is their meaning in ExFT. Let us proceed with short description of the split-form DFT mainly following \cite{Hohm:2013nja}.

\section{Split form of  DFT}
\label{sec:split}

 Start with generalised Lie derivative in O(10,10) theory which on the generalised vielbein takes the following form
\begin{equation}
\mL_V \mE_{\mM}{}^{\mA}=V^\mN \dt_\mN \mE_\mM{}^{\mA}+ \mE_\mN{}^\mA \dt_\mM V^\mN-\mE_\mN{}^\mA \dt^\mN V_\mM.
\end{equation}
We decompose 20 coordinates $\XX^\mM$ in two sets of $2D$ and $2d$ coordinates denoted by $\XX^{\hM}$ and $\XX^M$ respectively. The former will then trivially decompose into the conventional space-time coordinates $x^\m$ and their duals which will trivially drop from the picture to reproduce the proper section constraint for the resulting $O(d,d)$ theory. 

The O(10,10) invariant tensor is decomposed as follows
\begin{equation}
\h_{\bM\bN}=
\begin{bmatrix}
 \h_{\hM \hN} & 0\\
 0 & \h_{MN}
\end{bmatrix}.
\end{equation}
This corresponds to the less conventional but more convenient for our purposes choice of the invariant tensor
\begin{equation}
\h_{\mM\mN}=
\begin{bmatrix}
0 & {\bf 1}_{D\times D} & 0 & 0 \\
{\bf 1}_{D\times D} & 0 & 0& 0 \\
0 & 0 & 0 & {\bf 1}_{d\times d} \\
0 & 0&  {\bf 1}_{d\times d} & 0
\end{bmatrix}.
\end{equation}
The section condition remains the same
\begin{equation}
\begin{aligned}
\h^{\mM \mN}\dt_\mM \otimes \dt_\mN&=\h^{\hM \hN}\dt_{\hM}\otimes \dt_{\hN}+\h^{MN}\dt_M \otimes \dt_N\\
&=\tdt^m\otimes \dt_m+\tdt^\m \otimes \dt_\m.
\end{aligned}
\end{equation}
In what follows we will always assume $\tdt^\m=0$ and hence the full O(10,10) section constraint drops into the section constraint of the split $D+(d+d)$ DFT.

To decompose the fields we impose the standard split ansatz for the metric and the 2-form field 
\begin{equation}
\begin{aligned}
G_{\m\n}&=g_{\m\n}-A_{\m}{}^{m}A_\n{}^n g_{mn}, && B_{\m\n}=b_{\m\n}-2 b_{m[\m}A_{\n]}{}^m+A_{\m}{}^{m}A_\n{}^n b_{mn},\\
G_{\m m}&=A_\m{}^n g_{mn},&& B_{\m m}=b_{\m m}+A_\m{}^n b_{nm}.\\
G_{mn}&=g_{mn}
, && 
B_{mn}=b_{mn}.
\end{aligned}
\end{equation}
This implies that the full generalised vielbein is written in the following block form
\begin{equation}
\hat{E}_{\hM}{}^{\hA}=
\left[
\begin{array}{c:c|c:c}
e_\m{}^\a & A_\m^m e_m^a & -b_{\m\r}e^\r_\b-\fr12 \mA_\m{}^M\mA_{\r M}e^\r_\b  & (-b_{\m p}-A_\m{}^qb_{qp})e^p_b\\
\hdashline
0 & e_m^a & b_{\r m}e^\r_\b & -b_{mp}e^p_b \\
\hline 
0  & 0 & e_\b^\n & 0 \\
\hdashline
0  & 0 & -A_\r{}^n e^\r_\b & e^n_b
\end{array}
\right].
\end{equation}
Rearranging rows and columns in the same way as for the tensor $\h_{\hM\hN}$ we arrive at
\begin{equation}
\begin{aligned}
\hat{E}_{\hM}{}^{\hA}&=
\left[
\begin{array}{c:c|c:c}
e_\m{}^\a & -b_{\m\r}e^\r_\b-\fr12 \mA_\m{}^M\mA_{\r M}e^\r_\b & A_\m^m e_m^a & (-b_{\m p}-A_\m{}^qb_{qp})e^p_b\\
\hdashline 
0 & e_\b^\n & 0 & 0 \\
\hline
0 & b_{\r m}e^\r_\b & e_m^a & -b_{mp}e^p_b \\
\hdashline
0 & -A_\r{}^n e^\r_\b & 0 & e^n_b
\end{array}
\right]\\
&=
\left[
\begin{array}{c:c|c}
e_\m{}^\a & -b_{\m\r}e^\r_\b-\fr12 \mA_\m{}^M\mA_{\r M}e^\r_\b & \mA_\m{}^N E_N{}^A\\
\hdashline 
0 & e_\b^\n & 0  \\
\hline
0 & -\mA_{\r M}e^\r_\b & E_M{}^A 
\end{array}
\right].
\end{aligned}
\end{equation}
For the inverse vielbein one has
\begin{equation}
\begin{aligned}
(\hat{E}^{-1})_{\hA}{}^{\hM}
&=
\left[
\begin{array}{c:c|c}
e_\a{}^\m & -b_{\n\r}e^\r_\a-\fr12 \mA_\n{}^N\mA_{\r N}e^\r_\a & -\mA_{\r}{}^{ M}e^\r_\a\\
\hdashline 
0 & e_\n^\b & 0  \\
\hline
0 & \mA_{\n N} E^N{}_A & E_A{}^M 
\end{array}
\right].
\end{aligned}
\end{equation}

NS fluxes are encoded in the generalised torsion of DFT $\mF_{\mM\mN\mK}$ which is also decomposed under the split. 
For the generalised vielbein above the following components of the generalised flux vanish identically
\begin{equation}
\begin{aligned}
\mF_{\a}{}^{\b\g}, && \mF^{\a\b\g}, && \mF^{\a\b A}, && \mF^{\a A B}.
\end{aligned}
\end{equation}
For the rest one has
\begin{equation}
\begin{aligned}
\mF_{\mu \n \r}& =  3\big({D}_{[\mu}b_{\nu \rho]} + {\mA}_{[\mu}{}^{M} {\dt}_{\nu}{{\mA}_{\rho] M}}-  {\mA}_{[\mu}{}^{M} {\mA}_{\nu}{}^{N} {\partial}_{N}{\mA}_{\rho] M} \big), \\
&= 3\Big({D}_{[\mu}b_{\nu \rho]} + {\mA}_{[\mu}{}^{M} {\dt}_{\nu}{{\mA}_{\rho] M}}- \fr13 {\mA}_{[\mu}{}^{M} [{\mA}_{\nu},{\mA}_{\rho]}]_{ M} \Big),\\
\mF_{\m \n}{}^\r&=2 e_\a{}^\r D_{[\m} e_{\n]}{}^\a, \\
\mF_{\m\n}{}^{ M} &= 2\dt_{[\m}\mA_{\n]}{}^M-[A_\m,A_\n]_{\mc C}{}^M,\\
\mF_{M \a}{}^{\b}&= 3e_\m{}^\b \dt_M e_\a{}^\m=-\mF_{M}{}^\b{}_\a,\\
\mF_{\mu M N} &= 6 {\partial}_{[M}{\mA}_{\mu N]}-3{E}_{[N}{}^{A} \dt_{\mu}{E}_{A M]}  +3  \mA_{\mu}\,^{K}{E}_{[N}{}^{A}  {\partial}_{K}{E}_{A M]} \\
&=-3{E}_{[N}{}^{A} D_{\mu}{E}_{A M]}\\
\mF_{ABC} &=-3 E_{[A}{}^ME_B{}^N\dt_M E_{C] N}.
\end{aligned}
\end{equation}
where $[\; ,\,]_{\mc C}$ is the generalised Lie bracket and we define $D_{\mu}=\dt_\m-\mL_{A_\m}$. The latter is the standard covariant derivative along ``external'' coordinates $\{x^\m\}$ of Exceptional Field Theory (see \cite{Hohm:2013jma} and further works) which is needed to keep the theory in the split-form covariant. 

The field content of the split-form DFT can be summarized as follows
\begin{equation}
\begin{aligned}
& g_{\m\n}, && b_{\m\n}, && A_\m{}^M, && \mH_{MN}, && d.
\end{aligned}
\end{equation}
Structure of the theory is the same as that of Exceptional Field Theory and hence the result obtained here can be expanded to the fields of 11-dimensional supergravity. In particular, one is interested in tadpole cancellation conditions coming from exotic branes of M-theory. For that one analyses Bianchi identities of fluxes of split-form DFT listed above.

\section{Bianchi identities and sources}
\label{sec:bianchi}

The standard procedure when constructing the split-form DFT is to impose the following Bianchi identities 
\begin{equation}
\label{known_BIs}
\begin{aligned}
\dt_{[M}\mF_{NKL]}+\fr34\mF_{MN}{}^P\mF_{KL]P}&=T_{MNKL},\\
\DD_{[\m}\mH_{\n\r\s]}-\fr34 \mF_{[\m\n}{}^M\mF_{\r\s]M}&=T_{\m\n\r\s},\\
\DD_{[\m}\mF_{\n\r] M}+\dt_M \mH_{\m\n\r}&=T_{\m\n\r}{}_{M},
\end{aligned}
\end{equation}
where the first line presents in the full theory and the others describe interaction of the potentials \eqref{pot_D} with NS 5-branes of different orientation \cite{Blair:2017hhy}. To cover the full set of potentials one would expect to have in addition Bianchi identities of the form
\begin{equation}
\begin{aligned}
\DD_{[\m}\F_{\n]MN}+\dt_{[M}\mF_{\m\n N]}+\dots&=T_{\m\n}{}_{MN},\\
\DD_{\m}\mF_{MNK}+\dt_{[M}\F_{\m NK]}+\dots &=T_{\m}{}_{MNK},\\
\vdots
\end{aligned}
\end{equation}
with fields $\F_{\m MN}$, which contain the components $H_{\m mn}$ of the full Kalb-Ramond field strength in decomposition $10=D+d$. 

The most straightforward way to derive such identities is to start with the O(10,10) theory with the only Bianchi identities \eqref{BIO1010}, and to reduce it into the $D+(d+d)$ theory. The reduction will give all Bianchi identities and defined the corresponding fields. Hence, we start with the full covariant Bianchi identities of the $O(10,10)$ DFT
\begin{equation}
\mZ_{\mM\mN\mK\mL}=\dt_{[\mM}\mF_{\mN\mK\mL]}-\fr34 \mF_{{\mc P}[\mM\mN}\mF^{\mc P}{}_{\mK\mL]}
\end{equation}
As before for simplicity we consider DFT in the B-frame which implies that the following fluxes vanish
\begin{equation}
\begin{aligned}
\mF_{\a}{}^{\b\g}, && \mF^{\a\b\g}, && \mF^{\a\b A}, && \mF^{\a A B}.
\end{aligned}
\end{equation}
Decomposing the full Bianchi identities  we obtain
\begin{equation}
\label{BIsplit}
\begin{aligned}
\mZ_{A B C D}=&\ {E}_{[A}\,^{M} {\partial}_{M}{{\mF}_{B C D]}}\,  - \frac{3}{4} {\mF}^{E}\,_{[A B} {\mF}_{C D ]  E},\\
\mZ_{\a A B C}=&\ e_{\a}^{\mu} {D}_{\mu}{{\mF}_{A B C}}  - 3\, {E}_{[A}\,^{M} {\partial}_{M}{{\mF}_{\alpha B C]}}  - 3 {\mF}^{D\, }\,_{[A B} {\mF}_{\alpha C] D\, } + 3\, {\mF}_{\beta [A B} {\mF}^{\beta}\,_{C] \alpha} \\ 
\mZ_{\alpha \beta A B}=&\ 3e_{\a}\,^{\mu} {D}_{\mu}{{\mF}_{\beta A B}}+ 3\, {E}_{A}\,^{M} {\partial}_{M}{{\mF}_{\alpha \beta B}}\,   - \frac{3}{2}\, {\mF}^{C}\,_{A B} {\mF}_{\alpha \beta C} - \frac{3}{2}\, {\mF}_{\gamma A B} {\mF}_{\alpha \beta}\,^{\gamma}\\
& - 3\, {\mF}^{C}\,_{A \alpha} {\mF}_{\beta B C} + 6\, {\mF}^{\gamma}\,_{A \alpha} {\mF}_{\beta \gamma B},\\
\mZ_{A B \alpha }{}^{\beta}=&\  6\, {E}_{[A}\,^{M} {\partial}_{M}{\mF_{B] \alpha}{}^{\beta}}\,  - 3{\mF}_{C \alpha}{}^{\beta} {\mF}^{C}\,_{A B} - 6\, {\mF}_{[B \gamma}{}^{\beta} {\mF}_{A] \alpha}{}^{\gamma},\\
\mZ_{\alpha \beta \gamma A}=&\   3\, e_{[\alpha}\,^{\mu} {D}_{\mu}{{\mF}_{\beta \gamma] A}}\, - {E}_{A}\,^{M} {\partial}_{M}{{\mF}_{\alpha \beta \gamma}}  + 3 {\mF}^{B}\,_{A [\alpha} {\mF}_{\beta \gamma] B}\\
& + {3}\, {\mF}\,_{A [\alpha}{}^{\delta} {\mF}_{\beta \gamma] \delta} -{3}\, {\mF}_{A \delta [\alpha } {\mF}_{\beta \gamma]}\,^{\delta},\\
{\mZ}_{\alpha \beta A}{}^{\gamma}=&\  - 6\, e_{\alpha}\,^{\mu} {D}_{\mu}{{\mF}^{\gamma}\,_{A \beta}}+3\, {E}_{A}\,^{M} {\partial}_{M}{{\mF}_{\alpha \beta}\,^{\gamma}}\, + 6\, {\mF}^B{}_{ [\alpha}{}^{\gamma} {\mF}_{\b] B A} \\
&+ 6\, {\mF}_{A [\a}{}^{\delta} {\mF}_{\b] \delta}\,^{\gamma}  + 3\, {\mF}_{A \delta}{}^{\gamma} {\mF}_{\alpha \beta}\,^{\delta},\\
{\mZ}_{\alpha \beta \gamma \delta}=&\  e_{[\alpha}\,^{\mu} {D}_{\mu}{{\mF}_{\beta \gamma \delta]}}\,  - \frac{3}{4}\, {\mF}^{A}\,_{[\alpha \beta} {\mF}_{\gamma \delta ] A} - \frac{3}{4}\, {\mF}_{[\alpha \beta}\,^{\epsilon} {\mF}_{\gamma \delta] \epsilon} - \frac{3}{4}\, {\mF}_{\e [\alpha \beta } {\mF}_{\gamma \delta]}\,^{\epsilon},\\
\mZ_{\alpha \beta \gamma}{}^{\delta} =&\  3\, e_{\alpha}\,^{\mu} {D}_{\mu}{{\mF}_{\beta \gamma}\,^{\delta}}\,  - {3}\, {\mF}^{\delta}\,_{A \alpha} {\mF}^{A}\,_{\beta \gamma} + \frac{9}{4}\, {\mF}_{\alpha \beta}\,^{\epsilon} {\mF}_{\gamma \epsilon}\,^{\delta},\\
\mZ_{\alpha \beta}{}^{\gamma \delta} =&\  6\, {\mF}^{[\gamma A}\,_{[\alpha} {\mF}^{\delta]}\,_{A \beta]}.
\end{aligned}
\end{equation}
For constant fluxes BI's involving fluxes with only doubled indices are equivalent to quadratic constraints of $D$-dimensional half-maximal gauged supergravity.  

Based on the potentials which enter the covariant Wess-Zumino action for NS 5-branes as constructed in \cite{Bergshoeff:2011zk} one concludes that in the 10-dimensional space split as $10=D+d$ these potentials are sourced in the corresponding $O(d,d)$ theory by differently oriented branes \cite{Blair:2017hhy}.  Consider for definiteness the case $D=6$ and list all options for the DFT monopole
\begin{equation}
\begin{aligned}
& 0 && 1 && 2 && 3 && 4 && 5 && | && 6 && 7 && 8 && 9 && &&\\
&\tm&&\tm&&   &&   &&   &&   && | &&\tm&&\tm&&\tm&&\tm&& D_{(2)}, && \mF_{(3)} \\
&\tm&&\tm&&\tm&&   &&   &&   && | &&   &&\tm&&\tm&&\tm&& D_{(3), M}, && \mF_{(2)}{}^M\\
&\tm&&\tm&&\tm&&\tm&&   &&   && | &&   &&   &&\tm&&\tm&& D_{(4), MN},&& \mF_{(1)}{}^{MN}\\
&\tm&&\tm&&\tm&&\tm&&\tm&&   && | &&   &&   &&   &&\tm&& D_{(5), MNK},&& \mF_{(0),MNK}\\
&\tm&&\tm&&\tm&&\tm&&\tm&&\tm&& | &&   &&   &&   &&   && D_{(6), MNKL}, && \mF_{(0),MNK}
\end{aligned}
\end{equation}
where $\times$ denotes the worldvolume directions, empty space denotes transverse directions none of which is the Taub-NUT direction of the monopole. The directions $\{6,7,8,9\}$ are doubled. The corresponding mangetic gauge potentials represented by a $p$-form $D_{(p)M_1\dots M_1}$ with $q$ antisymmetrised $O(d,d)$ indices are listed in the first column on the RHS. The second column contains the corresponding field strengths all of which but the last are just deRahm differential of the $p$-form gauge potential. 

The top form $D_{(0)MNKL}$ cannot have field strength of such form, however in \cite{Bergshoeff:2016ncb} it has been show to be
\begin{equation}
\mF_{(0)MNK}=\dt^LD_{(0)LMNK}
\end{equation}
at the linearized level. Indeed, the Bianchi identity for the flux $\mF_{MNK}$ can be encoded by the additional term in the action for the brane
\begin{equation}
\D S= \int \big(\dt_{M}\mF_{NKL}-\fr34\mF_{PMN}\mF_{QKL}\h^{PQ}\big)D_{(6)}^{MNKL}.
\end{equation}
Hence, for each Bianchi identity one is able to define the corresponding magnetic potential. This procedure produces a number of dual potential $p$-forms in $D$ external dimensions which transform in tensor representations of the T-duality group $O(d,d)$
\begin{equation}
\begin{aligned}
& D_{(D-4)} && \mZ_{\m\n\r\s} && \\
& D_{(D-3), M} && \mZ_{\m\n\r M} && \\
& D_{(D-2), MN} && \mZ_{\m\n MN} && D_{(D-2)} && \mZ_{\m\n\r}{}^\r \\
& D_{(D-1), MNK} && \mZ_{\m MNK} && D_{(D-1),M} && \mZ_{\m \r M}{}^\r \\
& D_{(D), MNKL} && \mZ_{MNKL}&& D_{(D),MN} && \mZ_{MN\r}{}^\r  && D_{(D)} && \mZ_{\m\n}{}^{\m\n}, 
\end{aligned}
\end{equation}
where for the flux $\mF_{\m\n}{}^\r$ we provide only the trace part since its traceless part corresponds to the standard dual graviton. The first column above gives the same magnetic potentials as listed before. The second column contains additional potentials which have shown to be sourced by NS 5-branes in \cite{Bergshoeff:2011zk}. In the standard picture these are non-dynamical however these seem to be necessary to ensure gauge invariance of the  Wess-Zumino action. These correspond to roots of zero length in decomposition of the fundamental representation of $E_{11}$ under the subalgebra $O(d,d)$.

Finally, the last column contains the potential $D_{(D)}$ which must correspond to the Bianchi identity $\mZ_{\m\n}{}^{\m\n}$, and which has also been observed among the gauge potentials interacting to non-standard branes. However, since the corresponding Bianchi identity does not contain derivative it is not completely clear how to define field strength for such gauge potential. This corresponds to roots of negative length squared in the decomposition.

\section{Discussion}
\label{sec:m}

Since T-duality always doubles the number of coordinates one is able to develop a fully T-duality-covariant theory for the 10-dimensional supergravity which is a theory on the 10+10-dimensional doubled space. The same does not seem possible for U-duality since already for 2-dimensional supergravity the U-duality group is $E_9$ which is an infinite-dimensional affine algebra. The full 11-dimensional theory would then have $E_{11}$ as the local symmetry group, whose fundamental representation is infinite and hence infinite is the number of coordinates and fields. As the outcome one is not able to write Bianchi identities for the full $E_{11}$ ExFT simply as \eqref{BIO1010}. For this reason of interest is the split-DFT construction which has the same structure as ExFTs and which allows to make certain general conclusions about Bianchi identities of such theories and to construct magnetic gauge potentials  interacting with non-standard branes of M-theory and the corresponding field strengths. 

The problem of defining gauge potentials for exotic branes of M-theory and their covariant field strengths has been observed in \cite{Bakhmatov:2017les} for $6^{(3,1)}$-brane. Background for such brane has been obtained as a U-dual of the KK6-monopole background inside the $SL(5)$ exceptional field theory. The $6^{(3,1)}$-brane and the KK6-monopole belong to the U-duality orbit interacting with the 7-form potentials in the $\bf 5$, $\bf 45$ and $\bf 70$ of SL(5). At the linearized level derivatives of these along the extended space that transform under the $\bf 10$ of SL(5) give field strengths in $\bf 10$, $\bf 15$ and $\bf 40$, which presumably correspond to the non-constant gaugings of D=7 maximal supergravity
\begin{equation}
\begin{aligned}
&\dt^{(\bf 10)} A_{\m_1\dots \m_7}{}^{(\bf 5)} &&\longrightarrow \mF^{(\bf 10)}+ \mF^{(\bf 40)} \Longleftrightarrow (\q_{mn}, Z^{mn,k}),\\
&\dt^{(\bf 10)} A_{\m_1\dots \m_7}{}^{(\bf 45)} &&\longrightarrow \mF^{(\bf 10)}+ \mF^{(\bf 15)} \Longleftrightarrow (\q_{mn}, Y_{mn}),\\
&\dt^{(\bf 10)} A_{\m_1\dots \m_7}{}^{(\bf 70)} &&\longrightarrow \mF^{(\bf 10)}+ \mF^{(\bf 40)} \Longleftrightarrow (Z^{mn,k}).
\end{aligned}
\end{equation}
Bianchi identities for the SL(5) ExFT from which one would be able to derive the desired gauge potentials are not known explicitly. Bianchi identities of the split-form DFT as described above provide a guiding principle for constructing these in ExFT.

Indeed, consider the non-zero fluxes of the split-DFT in the B-frame
\begin{equation}
\begin{aligned}
&\mF_{\m\n\r}, && \mF_{\m\n}{}^\r, && \mF_{\m\n}{}^M, && \mF_{ABC},\\
& \mF_{M\a}{}^\b, && \mF_{\m}{}^{MN}.
\end{aligned}
\end{equation}
Since all these components came from the generalised flux $\mF_{\mM\mN\mK}$ of the full O(10,10) DFT, these can be thought of as fluxes or field strengths in the $D+(d+d)$ DFT as well. Expressions in the first line above indeed are usually understood as field strengths for the fields $B_{\m\n}$, $e_\m{}^\a$, $A_\m{}^M$ and $E_M{}^A$ respectively. Linearization must be there for the field strength of the dual graviton, both in the normal and doubled directions. The second line above contains spin-connections for the local Lorenz group $\mF_{M\a}{}^\b$ and the local group of the generalised diffeomorphisms $\mF_{\m}{}^{MN}$. From this one concludes that the full set of Bianchi identities of exceptional field theory must include connections and their derivatives understood as proper field strengths. Some progress in this direction has been made in \cite{Chatzistavrakidis:2019seu} where dynamical fluxes of the scalar sector of the SL(5) ExFT has been analyzed by considering Courant brackets of the Type II fluxes. The ideas described above provide extension of their results to the full ExFT including fluxes of the tensor sector.

The second observation is that the non-derivative parts of a subset of the Bianchi identities \eqref{BIsplit} has the same form as the quadratic constraints of half-maximal supergravity. Based on this one immediately writes Bianchi identities as the quadratic constraints of the corresponding maximal supergravity supplemented by terms containing derivatives of gaugings. Hence, one turns to dynamical non-geometric fluxes and understands them as proper field strengths.

\section*{Acknowledgements}   This work was supported by the Russian state grant Goszadanie 3.9904.2017/8.9, by the Foundation for the Advancement of Theoretical Physics and Mathematics ``BASIS'' and partially by the Alexander von Humboldt return fellowship.

\providecommand{\href}[2]{#2}\begingroup\raggedright\endgroup

\end{document}